\documentclass{iau_FM}
\usepackage{graphicx}
\usepackage{natbib}

\title[High-redshift radio galaxies] 
{High-redshift radio galaxies \\ at low radio frequencies} 

\author[A. Saxena \& H. J. A. R\"ottgering]   
{A. Saxena$^1$, \and H. J. A. R\"ottgering$^1$
}

\affiliation{$^1$ Leiden Observatory, Leiden University, \\
P.O. Box 9513, 2300 RA \\
Leiden, The Netherlands \\ email: {\tt saxena@strw.leidenuniv.nl} \\[\affilskip]
}

\pubyear{2018}
\jname{Astronomy in Focus, Volume 1} 
\editors{Volker Beckmann \& Claudio Ricci, eds.}
\begin{document}
\maketitle

\begin{abstract}
High-redshift radio galaxies (HzRGs) are some of the rarest objects in the Universe. They are often found to be the most massive galaxies observed at any epoch and are known to harbour active supermassive black holes that give rise to powerful relativistic jets. Finding such galaxies at high redshifts can shed light on the processes that shaped the most massive galaxies very early in the Universe. We have started a new campaign to identify and follow-up promising radio sources selected at 150 MHz in a bid to identify the most distant radio galaxies and study their properties, both intrinsic and environmental. Here we describe the progress of our campaign so far, highlighting in particular the discovery of the most distant radio galaxy known till date, at $z=5.72$.


\end{abstract}

\firstsection 
\section{Introduction}

Powerful radio galaxies are robust beacons of the most massive galaxies at any epoch. High-redshift radio galaxies (HzRGs) in particular are thought to be the progenitors of the massive elliptical galaxies that are observed in the local Universe. HzRGs contain an old stellar population \citep{bes98}, large amounts of dust and gas, are seen to be forming stars intensively \citep{sey08} and are often located in the centre of clusters and proto-clusters of galaxies \citep{ven02}. Studying their underlying stellar populations and their environment can shed light into the assembly and evolution of some of the most massive structures in the early Universe, including the evolution of the large scale structure of the Universe. \citet{mil08} give an extensive review about the properties of distant radio galaxies and their environments. Since HzRGs host large supermassive black holes (SMBHs) in their centres that power the bright radio synchrotron emission, detailed studies of radio galaxies at the highest redshifts can help test and constrain SMBH growth models and set estimates on the seed SMBH masses in the very early Universe. 

Bright radio galaxies at $z>6$, into the Epoch of Reionisation (EoR), are particularly interesting as they could in principle be used to study the process of reionisation in detail, constraining which remains one of the most important goals of modern cosmology. The redshifted 21cm (1.4 GHz) hyper-fine transition line of neutral hydrogen falls in the low radio frequency regime ($\nu < 200$ MHz) and can be observed as absorption in the spectra of a $z>6$ radio galaxy \citep{car02}. Such absorption signals from patches of neutral hydrogen could in principle be observed by current and next-generation radio telescopes such as the Giant Meterwave Radio Telescope (GMRT), the Low Frequency Array (LOFAR), the Murchinson Widefield Array (MWA) and the Square Kilometer Array (SKA). Detection of even a single bright radio source at $z>6$ could have profound implications for cosmology.

Motivated by the challenge of finding a HzRG deep in the EoR, we have designed a programme to isolate candidate HzRGs from large area radio surveys that go deeper than before. The shortlisted candidates are then followed up spectroscopically and photometrically, yielding distance measurements and galaxy properties. In the following sections we describe in detail our sample selection and some initial results

\section{Sample selection}

The sample of HzRGs presented in this paper were initially selected at a frequency of 150 MHz, from the TIFR GMRT Sky Survey Alternative Data Release (TGSS ADR; \citealt{int17}) covering $\sim10000$ square degrees of overlap with the VLA FIRST survey at 1.4 GHz. To isolate promising HzRG candidates from thousands of radio sources, we employed the ultra-steep spectrum selection ($\alpha^{150}_{1400} < -1.3$, where $S_\nu \propto \nu^\alpha$), which has proven to be very efficient at selecting HzRGs from large surveys \citep{rot94}. Flux limits were introduced so that a completely new parameter space in flux density and spectral index was probed, with all our sources having $50<S_{150}<200$ mJy. We then introduced angular size restrictions, as radio galaxies at the highest redshifts are expected to be compact \citep{sax17} and ensured that none of the promising candidates had any counterparts in the available all-sky optical and infrared surveys. Further details about the sample selection can be found in \citet{sax18a}.

The strict selection criteria resulted in a small sample of 32 promising candidates out of $\sim60000$ radio sources with spectral index information from TGSS ADR and FIRST. These 32 candidates were first followed up with the VLA at 1.4 GHz in A-configuration, to obtain high-resolution positions that would enable blind spectroscopy and accurate host galaxy identification in deep optical/infrared imaging. The radio maps at 1.4 GHz are presented in \citet{sax18a}.

\section{Multi-wavelength follow-up}
\subsection{Redshifts}
\begin{figure}
	\centering
	\includegraphics[scale=0.39]{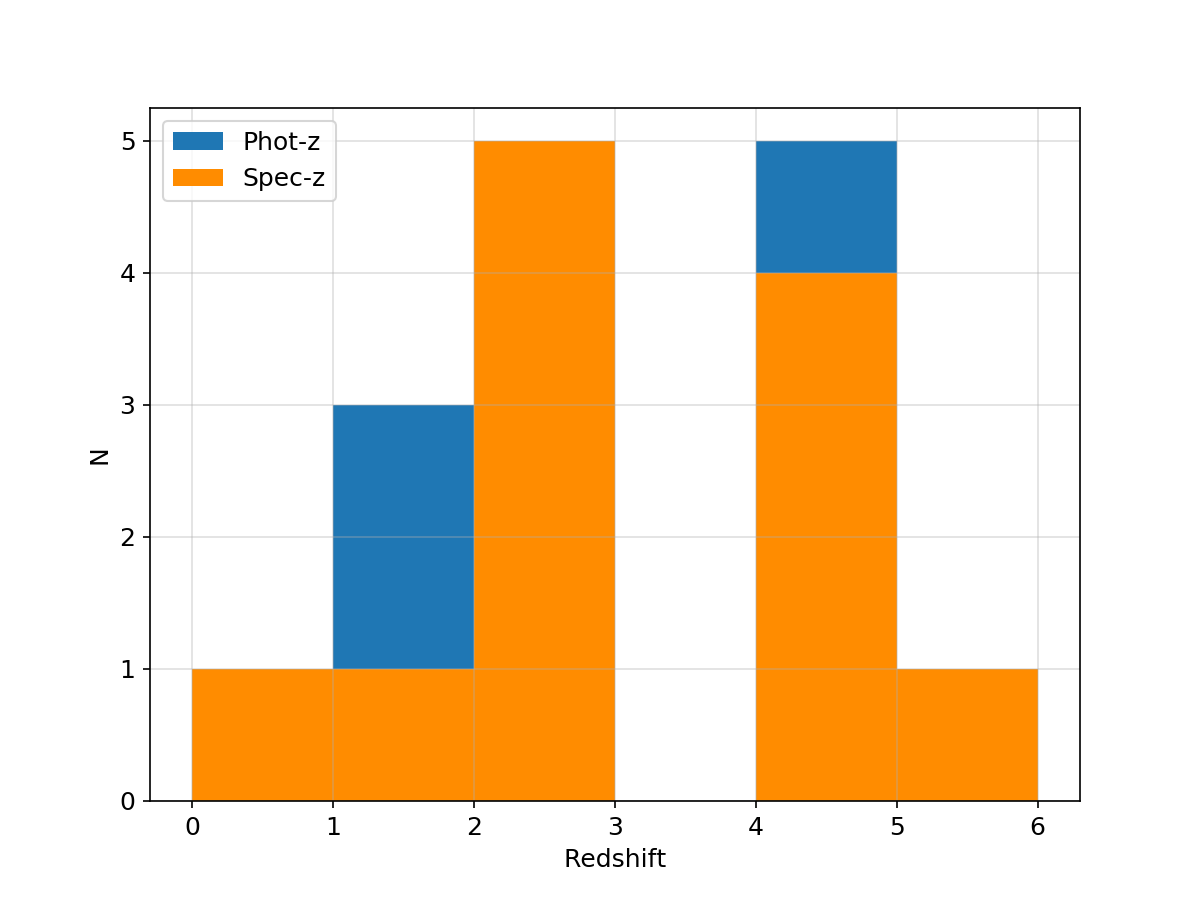}
	\caption{Distribution of spectroscopic and photometric redshift determined for the sample of radio galaxies presented in \citep{sax18a}.}
	\label{fig:redshift}
\end{figure}

A sub-sample of the 32 shortlisted sources was observed spectroscopically using GMOS on Gemini North, the William Herschel Telescope (WHT) and the Hobby-Eberly Telescope (HET). A few of these sources were also imaged at near infrared wavelengths in the $J$ and $K$ bands using the Large Binocular Telescope (LBT). Spectroscopic redshifts were obtained for 13 sources, and photometric redshifts were calculated for another two sources thanks to deeper optical and infrared data available in their field. We find redshifts in the range $0.57 < z < 5.72$ in our sample, with 5 newly discovered radio galaxies at $z>4$. There remain several radio sources from our initial sample of 32 that do not have any spectroscopic information yet and obtaining spectra for these sources is of the highest priority. The overall redshift distribution of the sample is shown in Figure \ref{fig:redshift} and the detailed analysis of the spectroscopic properties of the spectra is part of a manuscript in preparation. 

\subsection{Stellar masses}
Next, we calculate stellar masses for radio galaxies that have a redshift measurement and a $K$ band observation. We use simple stellar population models using \textsc{smpy}, a \textsc{python} package that enables creation of synthetic spectra in an easy and flexible manner. To model radio galaxy spectra, we use the assumptions outlined in \citet{sax18b} to create synthetic stellar population templates, that we then convolve with the broadband filter profiles at optical and infrared wavelengths. We then scale the synthetic photometry produced by these models, specifically in the $K$ band, to calculate stellar masses for each radio galaxy. The distribution of tracks on the $K-z$ parameter space and $K$ band magnitudes measured for sources in our sample will be presented in Saxena et al. (in prep). We find that our radio galaxies lie in the stellar mass range $10^{10.5} < M_\star < 11.7$ $M_\odot$, which is a bit lower than the stellar masses measured for powerful radio galaxies currently known across all epochs. Nonetheless, our radio galaxy stellar masses still appear to be representing the more massive end of the galaxy stellar mass function at all epochs, in line with previous studies, such as \citet{ove09}.

\section{Discovery of the highest redshift radio galaxy, at $z=5.72$}
\begin{figure}
	\centering
	\includegraphics[scale=0.18]{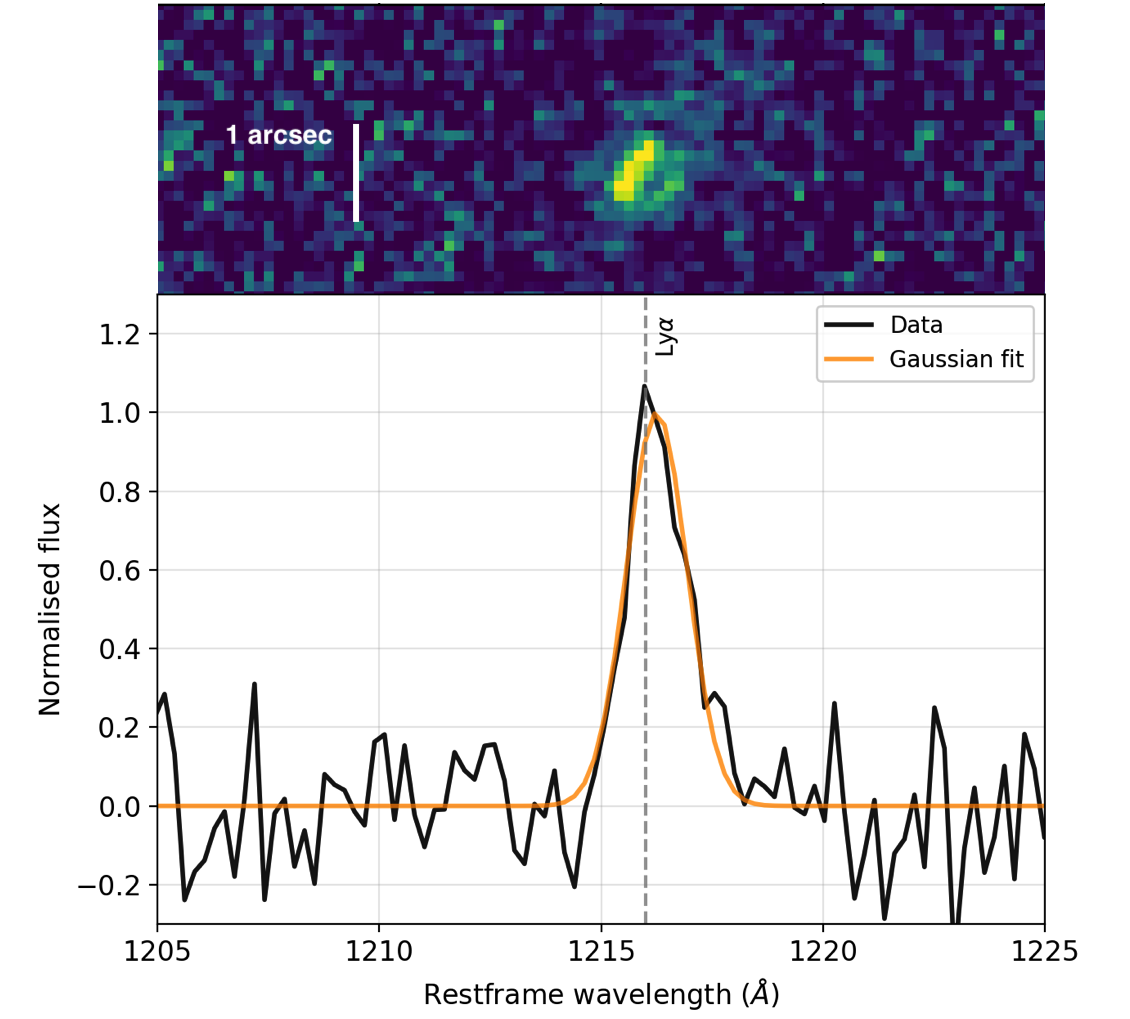}
	\caption{The rest-frame 1D spectrum (bottom) and 2D spectrum (top) for TGSS J1530+1049 at $z=5.72$ \citep{sax18b}. The FWHM is narrower than other HzRGs and is closer to that observed for `non-radio' LAEs at $z\sim5.7$.}
	\label{fig:spectrum}
\end{figure}

In this section we draw attention to the discovery of the highest redshift radio galaxy discovered to date, TGSS J1530+1049, which was initially part of our sample. The discovery of this galaxy was first reported in \citet{sax18b}. The redshift was determined using the Ly$\alpha$ emission line, which is slightly asymmetric, in line with what is seen for Lyman-alpha emitting galaxies (LAEs) at high-$z$ \citep{kas11}. Its faint $K$ band magnitude $K>24.3$ AB is also indicative of a high-redshift nature, owing to the $K-z$ relation that exists for radio galaxies \citep{wil03}. The Ly$\alpha$ line flux is $1.6 \times 10^{-17}$ erg s$^{-1}$ cm$^{-2}$ and the FWHM is $\sim370$ km s$^{-1}$, which is closer to that of normal LAEs at $z\sim5.7$ than other known $z>4$ radio galaxies. The line emission from TGSS J1530+1049 is shown in Figure \ref{fig:spectrum}.

TGSS J1530+1049 has a bright radio luminosity of $L_{150} = 10^{29.1}$ W Hz$^{-1}$ and the deconvolved radio size is measured to be $\sim3.5$ kpc, which is very compact. The relatively weak Ly$\alpha$, bright radio luminosity and a small radio size suggest that TGSS J1530+1049 is a radio galaxy that is at a very early stage in its lifetime. The stellar mass limit, under assumptions highlighted in \citet{sax18b} are $M_\star < 10^{10.5}$ $M_\odot$, which suggest that the host galaxy of this radio source is still in the process of assembling its stellar mass.

\section{Going forward}
Having successfully demonstrated that sensitive large area radio surveys can be used to hunt for radio galaxies at record distances, the big challenge facing the field now is to detect a radio galaxy deep in the EoR ($z>6$). Additionally, statistically significant samples of fainter radio galaxies across all redshifts can shed further light into the role of radio mode AGN feedback in shaping massive galaxies that we observe in the local Universe. Studying the underlying stellar populations and star formation histories for fainter, more `normal' radio galaxies would go a long way in gaining a complete picture about the formation and evolution of massive (radio) galaxies in the Universe.

\end{document}